\documentclass[useAMS,usenatbib,onecolumn]{mn2e}
\usepackage{epsfig,amssymb,amsmath,graphicx,enumitem}

\newcommand{\be}{\begin{equation}}
\newcommand{\ee}{\end{equation}}

\newcommand{\rstar}{R_\ast}
\newcommand{\mstar}{M_*}

\newcommand{\msun}{M_\odot}

\title[AXP 1E 2259$+$586 antiglitch and its internal magnetization]{Interpreting the AXP 1E 2259$+$586 antiglitch as a change in internal magnetization}
\author[A. Mastrano, A. G. Suvorov, and A. Melatos]{A. Mastrano\thanks{E-mail:
alpham@unimelb.edu.au}, A. G. Suvorov\thanks{E-mail: suvorova@student.unimelb.edu.au}, and A.
Melatos\thanks{E-mail: amelatos@unimelb.edu.au}\\School of Physics, University of Melbourne, Parkville VIC
3010, Australia}

\begin{document}

\date{Accepted ?. Received ?; in original form ?}

\pagerange{\pageref{firstpage}--\pageref{lastpage}} \pubyear{?}

\maketitle

\label{firstpage}

\begin{abstract}

\noindent{The sudden spin-down event (`anti-glitch') observed in AXP 1E 2259$+$586 on 2012 April 21 was arguably caused by a decay of its internal toroidal magnetic field component, which turns a stable prolate configuration into an unstable one. We refine previous models of this process by modelling the star's magnetic field self-consistently as a `twisted torus' configuration in non-barotropic equilibrium (which allows us to explore a greater range of equilibrium configurations). It is shown that, if the star's magnetic field is purely dipolar, the change in the toroidal field strength required to produce an anti-glitch of the observed size can be $\sim 10$ times larger than previously calculated. If the star has a quadrupolar magnetic field component, then an anti-glitch of similar magnitude can be produced via a decay of the quadrupole component, in addition to a decay of the toroidal component. We show that, if the quadrupole component decays, the minimum initial toroidal field strength and the change in toroidal field strength needed to produce the observed anti-glitch are lower than in the pure dipole twisted torus. In addition, we predict the maximum anti-glitch sizes, assuming that they are caused by a change in ellipticity, in four glitching magnetars and discuss the implications for energetics of accompanying X-ray bursts.}

\end{abstract}


\section{Introduction}

All neutron stars undergo electromagnetic braking \citep{lgs06}. In addition, some neutron stars are observed to undergo sudden, randomly occurring jumps in angular velocity of either sign. Many radio pulsars, as well as five magnetars,\footnote{A list of events is kept at http://www.physics.mcgill.ca/$\sim$pulsar/magnetar/main.html} exhibit sudden spin ups, known as `glitches' \citep{eetal11,yetal13,ok14,grs15}. One magnetar, the anomalous X-ray pulsar (AXP) 1E 2259$+$586, has been observed to experience a sudden spin \emph{down}, an `anti-glitch', accompanying an X-ray burst on 2012 April 21 \citep{aetal13b}. Recently, \citet{grs15} proposed that the anti-glitch was caused by a rearrangement of the internal magnetic field of the magnetar: long-term decay of the toroidal field component led to a change from a prolate to another less prolate mass distribution. In this paper, we present a refinement of the Garc\'{i}a \& Ranea-Sandoval model, by constructing a non-barotropic model star in equilibrium, whose density is perturbed by an imposed magnetic field. By considering a non-barotropic model, where pressure is not solely a function of density, we relax some constraints on the magnetic field configuration, allowing a wider range of astrophysical scenarios to be modelled.


Magnetars possess some of the strongest known, naturally occurring magnetic fields \citep{mpm15}. The external field strength of a neutron star is usually inferred from its spin down. The internal field, however, cannot be measured directly. Observations of bursts and giant flares \citep{i01,co11} and precession \citep{metal14} in magnetars have been interpreted to indicate that the internal field exceeds the external field by at least one order of magnitude. Numerical simulations favour a `twisted torus' magnetic configuration \citep{bn06,bs06}. Magnetic fields of these strengths deform the stellar mass distribution away from spherical symmetry.

Recently the magnetic deformation of non-barotropic neutron stars has been calculated. Gravitational wave observations (from which we can obtain ellipticity $\epsilon$) can thus be used to infer the internal magnetic structure of neutron stars in principle \citep{metal11,mlm13,msm15}. Because the stars are assumed to be non-barotropic [e.g., due to entropy or lepton fraction gradients \citep{rg92,r01,r09}], a greater range of MHD equilibria can be constructed and analysed, and one can easily construct good analytic approximations to the linked poloidal-toroidal twisted torus found in numerical simulations \citep{bn06,b09,aetal13,detal14}. Although magnetars rotate too slowly to be good gravitational-wave source candidates, the above models can be used to deduce the internal field strengths of fast-rotating newborn magnetars (age $\lesssim 10$ s), which are better candidates for a hypothetical gravitational wave detection from distances as far as the Virgo cluster \citep{ds07,dss09,metal11}.

The AXP 1E 2259$+$586 suffered an overall spin frequency change $\approx -5\times 10^{-7}$ Hz over $\sim 10^2$ d \citep{aetal13b}. According to \citet{aetal13b}, there are two possible explanations for this event: an anti-glitch with $\Delta\nu/\nu=-3.1(4)\times 10^{-7}$ followed by a glitch with $\Delta\nu/\nu=2.6(5)\times 10^{-7}$, or an anti-glitch with $\Delta\nu/\nu=-6.3(7)\times 10^{-7}$ followed by another anti-glitch with $\Delta\nu/\nu=-4.8(5)\times 10^{-7}$. Bayesian analysis by \citet{hetal14} showed that the second explanation, a double anti-glitch, fits the data better. \citet{grs15} interpreted the angular velocity drop in terms of a change in $\epsilon$. This interpretation is exciting because it offers a way to infer the internal field strength and structure of AXP 1E 2259$+$586 independent of other measurements of $\epsilon$, e.g., from future gravitational wave observations of similar objects at the beginning of their lives. However, \citet{grs15} used a uniform-density star as the unmagnetized background state, ignored the contribution of the poloidal field component to $\epsilon$, and assumed a uniformly distributed internal field (rather than some self-consistent, spatially varying equilibrium configuration). We relax these restrictions in three ways: we analyse a non-uniform parabolic density distribution [which is sufficient to approximate an $n=1$ polytropic star: see \citet{metal11} and Sec. 2.1 of this paper], a non-barotropic hydromagnetic equilibrium, and a linked poloidal-toroidal field structure. We calculate $\epsilon$ and explore how much the toroidal field component needs to change to generate the observed anti-glitch magnitude. We also explore the possibility that AXP 1E 2259$+$586 has a quadrupolar field component, which can contribute to the anti-glitch if it changes.

In Section 2, we recap the non-barotropic deformation calculations in the literature \citep{metal11,msm15} and apply them to AXP 1E 2259$+$586 for a dipolar twisted torus configuration. We investigate a range of initial and final states, that give the observed anti-glitch magnitude, comparing our results directly to those of \citet{grs15}. In Section 3, we construct a dipole-plus-quadrupole twisted torus, apply it to AXP 1E 2259$+$586, and repeat the test over a range of initial and final states. In Section 4, we calculate the energy released during these changes in field configuration and compare them to the observed outburst energy. We keep the dipole poloidal field strength $B_p=5.9\times 10^{13}$ G constant throughout this paper, corresponding to the spin-down-inferred value, as \citet{grs15} did. In other words, like they did, we assume that the frequency change is purely due to a decay of the internal toroidal field (Section 2) or the quadrupole component (Section 3), and the dipole poloidal field (which is responsible for overall spin down) remains unchanged before and after the glitch \citep{grs15}. In Section 4, we calculate the change in magnetic energy accompanying this process and compare it to the observed X-ray burst energy. In Section 5, we predict the maximum sizes of possible anti-glitches in other magnetars caused by this mechanism and their plausibility. In Section 6, we summarize our results, discuss briefly other possible causes of the anti-glitch, and discuss future work needed to refine the Garc\'{i}a \& Ranea-Sandoval anti-glitch model.

\section{Dipole twisted torus}

In section 2.1, we summarize the calculation of \citet{metal11}, which relates $\epsilon$ to the strength of the toroidal magnetic field component in a non-barotropic star with a twisted torus, whose poloidal component is purely dipolar. In Section 2.2, we apply the results to the AXP 1E 2259$+$586 anti-glitch.

\subsection{Hydromagnetic equilibrium}

The magnetic field is assumed to be axisymmetric. It is decomposed into its poloidal and toroidal components and expressed in dimensionless spherical polar coordinates ($r,\theta,\phi$) as \citep{c56,metal11,mm12,mlm13},

\be {\bf{B}}=B_0 [\eta_p \nabla\alpha(r,\theta)\times \nabla\phi + \eta_t \beta(\alpha)\nabla\phi],\ee
where $B_0$ parametrizes the overall strength of the field, $\eta_{p}$ and $\eta_t$ set the relative strengths of the poloidal and toroidal components respectively ($\eta_p= 1$ without loss of generality), $\alpha(r,\theta)$ is the poloidal magnetic stream function, and the function $\beta(\alpha)$ sets the toroidal field component. We consider separable stream functions of the form $\alpha(r,\theta)=f(r)g(\theta)$ in this paper; for a dipolar configuration, one has $\alpha=f(r) \sin^2\theta$. The radial part of the stream function, $f(r)$, is formally arbitrary, as long as it results in a field that is axisymmetric, continuous everywhere inside the star ($r<1$), continuous with a current-free dipolar external field at  the surface ($r=1$), finite everywhere, and whose current vanishes at $r=1$. For analytic simplicity, we choose the polynomial \citep{metal11}

\be f(r) = \frac{35}{8} \left(r^2-\frac{6r^4}{5}+\frac{3r^6}{7}\right).\ee
The function $\beta$ must be a function of $\alpha$ to ensure that the magnetic force has no azimuthal component, which cannot be balanced in magnetohydrostatic equilibrium given a field of the form (1) \citep{metal11,mlm13}. To conform to the numerical simulations of \citet{bn06} and \citet{b09}, we want the toroidal component to be confined to a circumstellar torus near the neutral curve, i.e., the circle where the poloidal field component vanishes, located at some radius $r=r_\textrm{N}$ and $\theta=\pi/2$ (with $r_\textrm{N}=0.79$ for the dipole twisted torus in Sections 2.1 and 2.2). Hence, in keeping with previous works, we choose

\be \beta(\alpha) =
\begin{cases}
(\alpha - 1)^2&\textrm{for }\alpha\geqslant 1,\\
0&\textrm{for }\alpha < 1.
\end{cases}
\ee

We treat the magnetic force as a perturbation on a background hydrostatic equilibrium and write the hydromagnetic force balance equation as

\be\frac{1}{\mu_0} (\nabla\times {\bf{B}})\times{\bf{B}}=\nabla\delta p +\delta\rho\nabla\Phi,\ee
to first order in $B^2/(\mu_0 p)$, in the Cowling approximation $(\delta\Phi = 0)$, where $p$ is the zeroth-order pressure, $\rho$ is the zeroth-order density, $\Phi$ is the gravitational potential, and $\delta p$, $\delta\rho$, $\delta\Phi$ are perturbations of the latter three quantities. \citet{y13} calculated $\epsilon$ for a pure dipole without taking the Cowling approximation. His values of $|\epsilon|$ are at most $\sim 2$ times those found by \citet{metal11} (depending on the choice of zeroth-order density profile). A thorough, full-perturbation calculation without the Cowling approximation is beyond the scope of this paper.

We do not assume a barotropic star, nor do we solve the Grad-Shafranov equation, so $\delta\rho$ does not have to be a function solely of $\delta p$, and therefore the equation of state imposes no restrictions on the field structure. Neutron star matter consists of multiple species which reach a stably stratified, hydromagnetic equilibrium within a few Alfv\'{e}n time-scales \citep{p92,rg92,r01}. This system is not in full chemical equilibrium, however; the relative abundances of the constituent particles change by weak nuclear interactions [time-scale $\sim 10^5 (T/10^8\textrm{ K})^{-6}$ yr] and diffusive processes [time-scale $\sim 10^9(B/10^{11}\textrm{ T})^{-2}(T/10^8\textrm{ K})^{-6}$ yr]. Between the Alfv\'{e}n and the weak nuclear time-scales \citep{hrv08}, the star is in a hydromagnetic equilibrium state, in which the composition is not determined solely by the density or pressure, and density and pressure do not correspond one-to-one \citep{metal11}. Since most magnetars are 1--10 kyr old (as inferred from spin and spin down), the non-barotropic assumption is valid.\footnote{Due to the strong inverse dependence of the diffusion and weak nuclear time-scales on temperature, and the rapid cooling of neutron stars over $\sim 10^2$ kyr, chemical equilibrium may never be reached in practice \citep{yls99,pr10,pr11,gjpr14}, so our non-barotropic assumption may be valid for more than just the relatively young magnetars.}

For the zeroth-order density, we choose a parabolic density profile

\be \rho(r) = \rho_\textrm{c}(1-r^2),\ee
where $\rho_\textrm{c}=15\mstar/(8\pi \rstar^3)$ is the density at origin, $\mstar$ is the stellar mass, and $\rstar$ is the stellar radius. This choice is merely for computational simplicity, but, as shown by \citet{metal11}, the resulting $\epsilon$ agrees within 5 per cent with that from a more realistic $n=1$ polytrope.

The stellar ellipticity is given by

\be \epsilon = \frac{I_{zz}-I_{xx}}{I_0},\ee
where $I_0$ is the moment of inertia of the unperturbed spherical star, the moment-of-inertia tensor is given by

\be I_{jk} = R^5_* \int_V \textrm{d}^3x[\rho(r)+\delta\rho(r,\theta)](r^2\delta_{jk}-x_j x_k),\ee
and the integral is taken over the volume of the star $(r\leqslant 1)$. The density perturbation $\delta\rho$ is calculated by taking the curl of both sides of equation (4) and matching the $\phi$-components:

\be \frac{\partial\delta\rho}{\partial\theta}=-\frac{r}{\mu_0 \rstar}\frac{\mathrm{d}r}{\mathrm{d}\Phi}\{\nabla\times[(\nabla\times{\bf{B}})\times{\bf{B}}]\}_\phi.\ee

For a dipolar twisted torus with $\alpha(r,\theta)=f(r)\sin^2\theta$, \citet{metal11} and \citet{mm12} derived the following formula relating $\epsilon$ to overall field strength, stellar mass, stellar radius, and the relative poloidal and toroidal field strengths:

\be \epsilon = 5.63\times 10^{-6} \left(\frac{B_p}{10^{15}\textrm{ G}}\right)^{2} \left(\frac{\rstar}{10^4\textrm{ m}}\right)^4 \left(\frac{\mstar}{1.4\textrm{ }\msun}\right)^{-2}\left(1-\frac{0.351}{\Lambda}\right).\ee
In equation (9), $B_p$ is the surface field strength at the pole (for a pure dipole, one has $B_p=2B_0$), $\Lambda$ is the ratio of internal poloidal field energy to total internal field energy, $\Lambda=0$ gives a purely toroidal configuration and $\Lambda=1$ gives a purely poloidal configuration.

\subsection{AXP 1E 2259$+$586 anti-glitch}

Armed with equation (9), we repeat the calculation of \citet{grs15}, who calculated the change in $\epsilon$ as a function of the anti-glitch's frequency change. Section 2 of the latter reference contains the result

\be \frac{\Delta\nu}{\nu}\approx \frac{2}{3}(\epsilon_\textrm{i}-\epsilon_\textrm{f}),\ee
where $\nu$ is the star's rotation frequency, $\Delta\nu$ is the change in frequency during the anti-glitch, and $\epsilon_\textrm{i}$ $(\epsilon_\textrm{f})$ is the initial (final) ellipticity. Equation (10) is derived by assuming that the star initially contains a mostly toroidal field (which induces $\epsilon_\textrm{i}$), which slowly decays by Ohmic diffusion and Hall drift \citep{vetal13}. Some critical strain is then reached, when the crust cracks in a sudden event (i.e., the outburst), and the star settles into a new field configuration with $\epsilon_\textrm{f}>\epsilon_\textrm{i}$. The change in frequency (i.e., the anti-glitch) arises from a change in moment of inertia (because angular momentum is conserved), which is then related to the change in $\epsilon$ to give equation (4) in \citet{grs15}, reproduced here as equation (10).

The measured frequency change associated with the anti-glitch is $-5 \times 10^{-7}$ Hz over $\sim 10^2$ d \citep{aetal13b}. To explain this, \citet{aetal13b} proposed that the event consisted of (1) an anti-glitch with $\Delta\nu/\nu=-3.1(4)\times 10^{-7}$ followed by a glitch with $\Delta\nu/\nu=2.6(5)\times 10^{-7}$, or (2) an anti-glitch with $\Delta\nu/\nu=-6.3(7)\times 10^{-7}$ and another anti-glitch with $\Delta\nu/\nu=-4.8(5)\times 10^{-7}$. The second explanation, a double anti-glitch, is statistically favoured \citep{hetal14}. We take the measured value $\Delta\nu/\nu=-6.3\times 10^{-7}$ [both for definiteness and to facilitate direct comparison with the results presented in Section 3 of \citet{grs15}; the calculations can easily be repeated for $\Delta\nu/\nu=-4.8\times 10^{-7}$] and solve for $\epsilon_\textrm{f}$ in equation (10) for various $\epsilon_\textrm{i}$. We present the results in Figs. 1 and 2, to be compared directly to Figs. 1 and 2 of \citet{grs15}. To facilitate comparison with \citet{grs15}, we express the toroidal field strength as $\langle B_t\rangle$, i.e., $B_t$ averaged over the volume of the torus. We hold $B_p=5.3\times 10^{13}$ G constant, corresponding to the value inferred from spin down, and allow $B_t$ to change.

\begin{figure}
\centerline{\epsfxsize=14cm\epsfbox{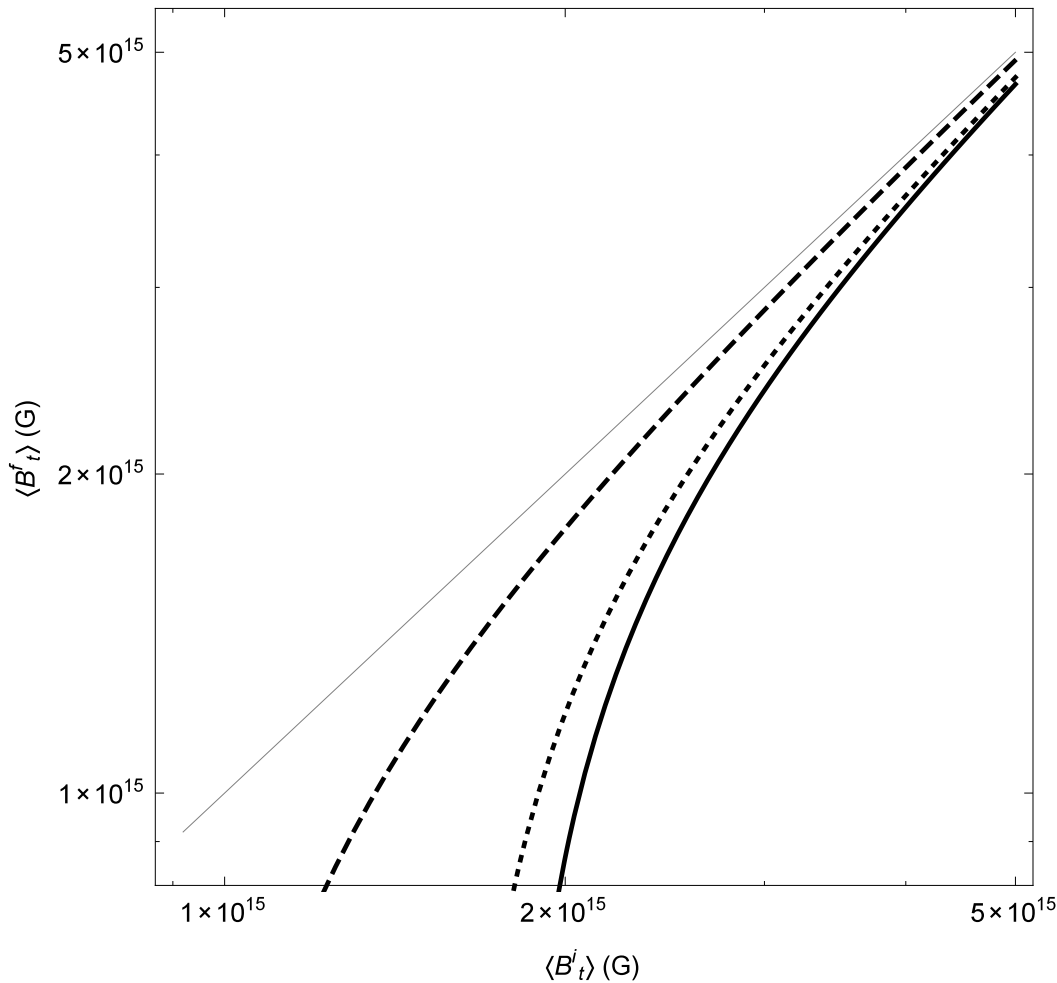}}
 \caption{Initial and final toroidal field combinations consistent with the observed anti-glitch in the AXP 1E 2259$+$586, i.e., solutions to equation (10) for $\Delta\nu/\nu=-6.3\times 10^{-7}$ [cf. Fig. 1 of \citet{grs15}] as a function of initial volume-averaged toroidal field strength $\langle B^i_t\rangle$, for surface polar field strength $B_p=5.9\times 10^{13}$ G (inferred from spin down) and three combinations of stellar masses and radii [chosen by \citet{grs15}]: $\mstar=1.4\msun$, $\rstar=10^4$ m (solid curve); $\mstar=1.4\msun$, $\rstar=1.4\times 10^4$ m (dashed curve); and $\mstar=1.8\msun$, $\rstar=1.2\times 10^4$ m (dotted curve). The thin solid curve is $\langle B^i_t\rangle = \langle B^f_t\rangle$ for reference.}
 \label{fig1grs}
\end{figure}

Figure \ref{fig1grs} shows that the toroidal field strength must decrease from its initial value to give an anti-glitch of the observed magnitude, as noted by \citet{grs15}, with each curve approaching a particular value of $\langle B^i_t\rangle$ as $\langle B^f_t\rangle \rightarrow 0$. This minimum allowed value for $\langle B^i_t\rangle$ is found by setting $\epsilon_\textrm{f}=\epsilon_\textrm{max}$ in equation (10), where the maximum ellipticity $\epsilon_\textrm{max}$ is obtained by setting $\Lambda=1$ (a purely poloidal configuration) in equation (9). However, we require higher initial $\langle B^i_t\rangle$, as well as larger changes in $\langle B_t\rangle$, than predicted by \citet{grs15}. For example, in order for a star with $\mstar=1.4$ $\msun$ and $\rstar=10^4$ m [solid curves in Fig. \ref{fig1grs} in this paper and in \citet{grs15}] to give $\Delta\nu/\nu=-6.3\times 10^{-7}$, the toroidal field must decrease from $\langle B^i_t\rangle=2\times 10^{15}$ G to $\langle B^f_t\rangle\approx 8.7\times 10^{14}$ G, whereas \citet{grs15} calculated $\langle B^f_t\rangle\approx 1.8\times 10^{15}$ G. This is because the poloidal field component tends to deform the star into an oblate shape; neglecting it overestimates the prolateness of the star for a given $\langle B_t\rangle$. This overestimate of $\epsilon$ means that, to obtain a given $|\epsilon_\textrm{i}-\epsilon_\textrm{f}|$, we need $\langle B_t\rangle$ to change more than predicted by \citet{grs15}.

\begin{figure}
\centerline{\epsfxsize=14cm\epsfbox{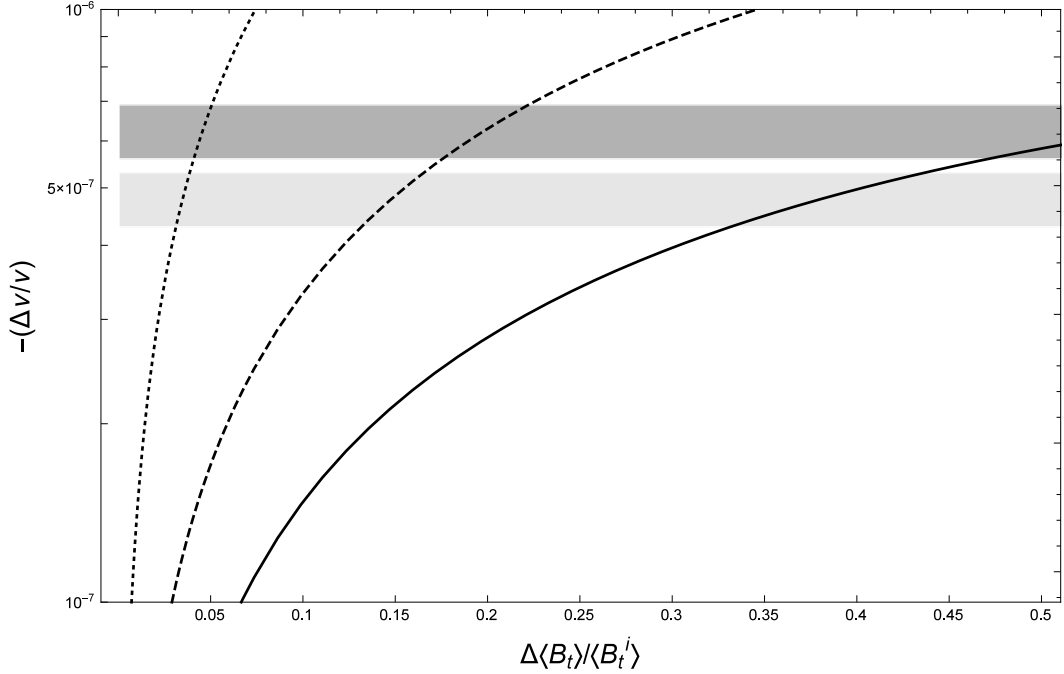}}
 \caption{The fractional change in frequency $\Delta\nu/\nu$ as a function of the relative change in the volume-averaged toroidal magnetic field strength $|\Delta\langle B_t\rangle|/\langle B^i_t\rangle$, where $\Delta\langle B_t\rangle = \langle B^i_t\rangle - \langle B^f_t\rangle$ [cf. Fig. 2 of \citet{grs15}], i.e., solutions to equation (10) compared to the observed anti-glitch in the AXP 1E 2259$+$586, for surface polar field strength $B_p=5.9\times 10^{13}$ G and three values of $\langle B^i_t\rangle$: $2\times 10^{15}$ G (solid curve); $3\times 10^{15}$ G (dashed curve); and $4.5\times 10^{15}$ G (dotted curve). The bands show $\Delta\nu/\nu$ for the two possible AXP 1E 2259$+$586 anti-glitch magnitudes inferred from the 2012 April data [taking the statistically favoured double anti-glitch interpretation \citep{hetal14}] and their corresponding error bars: $\Delta\nu/\nu=-6.3(7)\times 10^{-7}$ (the initial anti-glitch, dark grey) and $-4.8(5)\times 10^{-7}$ (the follow-up anti-glitch, light grey).}
 \label{fig2grs}
\end{figure}

The above behaviour is displayed from an alternative viewpoint in Fig. \ref{fig2grs}. We plot $-\Delta\nu/\nu$ as a function of the toroidal field strength change $\Delta\langle B_t\rangle/\langle B^i_t\rangle$, where $\Delta\langle B_t\rangle=\langle B^i_t\rangle - \langle B^f_t\rangle$, for three different values of $\langle B^i_t\rangle$, with $B_p=5.9\times 10^{13}$ G. In the same plot, the horizontal bands indicate $-\Delta\nu/\nu=6.3(7)\times 10^{-7}$ (upper band, dark grey) and $-\Delta\nu/\nu=4.8(5)\times 10^{-7}$ (lower band, light grey). Comparing the solid curve in Fig. \ref{fig2grs} with the dashed curve in Fig. 2 of \citet{grs15} (both representing $\langle B^i_t\rangle = 2\times 10^{15}$ G), we see that, to obtain $\Delta\nu/\nu = -6.3\times 10^{-7}$, our model requires $\Delta\langle B_t\rangle/\langle B^i_t\rangle\approx 50$ per cent, whereas \citet{grs15} calculated $\Delta\langle B_t\rangle/\langle B^i_t\rangle\approx 5$ per cent. Thus, by ignoring the poloidal field component, one overestimates $|\epsilon|$ in general and underestimates $\Delta\langle B_t\rangle/\langle B^i_t\rangle$ for a given $|\Delta\nu/\nu|$. For example, for $1.9\times 10^{15}\textrm{ G}<\langle B_t\rangle<4.9\times 10^{15}\textrm{ G}$ [corresponding to the minimum value needed for the anti-glitch and the expected limit of stability; see below and \citet{aetal13}] and $B_p=5.9\times 10^{13}$ G (inferred from spin down), we find $\epsilon/\epsilon_\textrm{GRS}\sim 0.15$, where $\epsilon$ is the ellipticity calculated using equation (9) and $\epsilon_\textrm{GRS}$ is the ellipticity calculated by \citet{grs15}.

The analytic calculation of \citet{aetal13} and the numerical simulation of \citet{b09} found that $10^{-3}\lesssim \Lambda\lesssim 0.8$ is required for a stable dipolar twisted torus. For $B_p=5.9\times 10^{13}$ G, this corresponds to $7.7\times 10^{13} \textrm{ G }\lesssim\langle B_t\rangle\lesssim 4.9\times 10^{15}$ G. A more stringent limit is set by crust cracking. \citet{hk09} conducted large-scale molecular dynamics simulations of Coulomb solids to represent a neutron star's crust. They found that the large density and pressure make the crust very strong and rigid, capable of supporting $|\epsilon|\lesssim 4\times 10^{-6}$ (assuming canonical neutron star mass and radius) before cracking suddenly in a collective manner (rather than yielding continuously). If we assume that the initial magnetic field does not deform the star sufficiently to crack the crust, then this limit translates to $\langle B^i_t\rangle\lesssim 3.7\times 10^{15}$ G in the case of AXP 1E 2259$+$586. The values of $\langle B_t\rangle$ plotted in Figs. \ref{fig1grs} and \ref{fig2grs} fall within these relatively generous limits.


\section{Dipole-plus-quadrupole twisted torus}

Neutron star magnetic fields are approximately dipolar at radio emission altitudes \citep{lm88,cm11} and in the outer magnetosphere, where high-energy emissions originate \citep{ry95,lom12}. However, some observations can be interpreted as signatures of higher-order multipoles close to the surface \citep{bm14}, e.g. cyclotron resonant scattering line energies of certain accretion-powered X-ray pulsars \citep{n05,pml14}, radio emissions from pulsars beyond the `death line' \citep{ymj99,cetal00,ml10}, the anomalous braking index of some radio pulsars \citep{bt10}, and the substructures found in some pulsar signals \citep{bmh15,p15}. While the dipole component of the magnetic field can be inferred from the observed spin-down rate, the putative higher-order multipoles contribute small corrections of order $\lesssim (2\pi\nu \rstar/c)^2$ to the torque and cannot be measured directly.

In this section, we outline the method used by \citet{msm15} to describe a self-consistent, linked dipole-plus-quadrupole-plus-toroidal field configuration and to relate the resulting $\epsilon$ to $\langle B_t^i\rangle$, $\langle B_t^f\rangle$, and the amplitude of the quadrupolar component.

\subsection{Hydromagnetic equilibrium}

If the poloidal field is a combination of dipole and quadrupole components, then the stream function $\alpha(r,\theta)$ can be written as a linear combination of the form

\be \alpha(r,\theta) = f_1(r) \sin^2\theta + \kappa f_2(r) \sin^2\theta\cos\theta,\ee
where $\kappa$ is a dimensionless parameter controlling the amount of quadrupole field present. In addition to the finiteness and continuity conditions discussed in Section 2.1, $f_1(r)$ and $f_2(r)$ must now be chosen such that the Lorentz force and the resulting density perturbation are continuous everywhere. This is not a trivial task and is best accomplished by moving to a coordinate system defined by the stream function, as described by \citet{msm15}. Suitable $f_{1,2}(r)$ are then found through trial and error. The calculation of $\epsilon$ itself is straightforward, using equations (6)--(8). Following \citet{msm15} for simplicity and consistency, we use

\be f_1(r)=\left(\frac{117}{32}-\sigma\right)r^4 - \left(\frac{65}{16}-3\sigma\right)r^8 + \left(\frac{45}{32}-3\sigma\right)r^{12}+\sigma r^{16},\ee
\be f_2(r)=\frac{1}{8}\left(35 r^4 - 42 r^8 + 15 r^{12}\right),\ee
where $\sigma$ is a dimensionless free parameter which controls the volume of the torus occupied by the toroidal field component. These $f_{1,2}(r)$ choices are not unique, but they are among the simplest possible polynomials, with the lowest possible order, that guarantee a well-defined field-aligned coordinate transformation.

The stability of dipole-plus-quadrupole configurations has not been calculated, either analytically [like \citet{aetal13}] or numerically [like \citet{bn06}]. As a consequence, some combinations of $\sigma$, $\kappa$, and $\langle B^{i,f}_t\rangle$ may be excluded in reality for stability reasons. We defer the numerical calculation (by evolving the field in a time-dependent magnetohydrodynamic simulation) to a future paper.

\subsection{AXP 1E 2259$+$586 anti-glitch}

In this section, we repeat the calculation of \citet{grs15}, except that we assume AXP 1E 2259$+$586 possesses some quadrupolar magnetic component, which partially decayed during the anti-glitch. In other words, we assume that the change in $\epsilon$ which led to the anti-glitch is due to a change in field geometry, in addition to a decay of the toroidal component. We hold $B_0=2.95\times 10^{13}$ G constant, so that the dipolar component's surface polar strength is $5.9\times 10^{13}$ G (we thereby assume that the measured spin down, from which a polar field strength of $5.9\times 10^{13}$ G is inferred, is due solely to the dipole component). We also set $\sigma=-5$ (for the sake of definiteness), and allow $\kappa$ and $\langle B_t\rangle$ to change.

\begin{figure}
\centerline{\epsfxsize=13cm\epsfbox{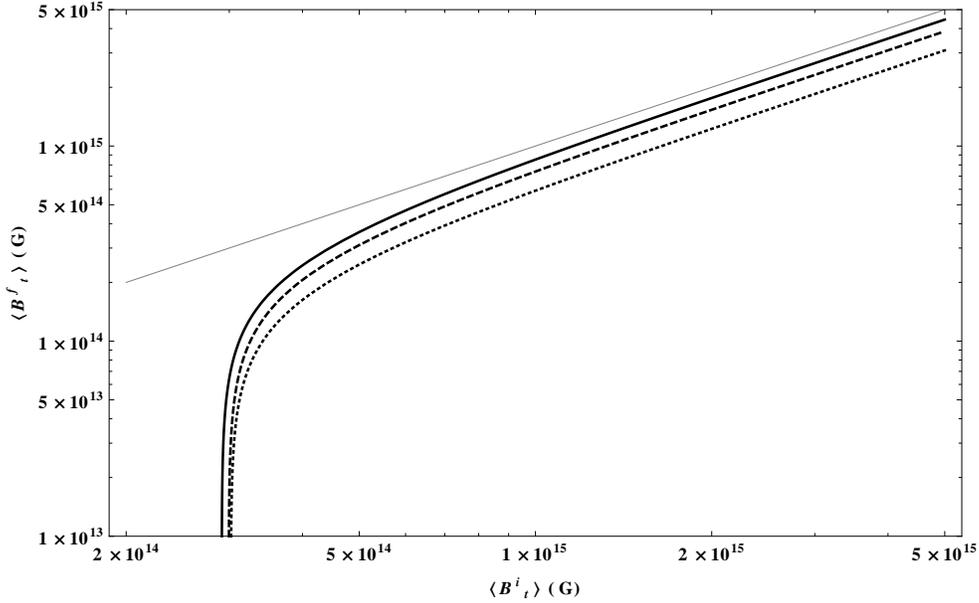}}
 \caption{Initial and final toroidal field combinations consistent with the observed anti-glitch in the AXP 1E 2259$+$586, i.e., solutions to equation (10) for $\Delta\nu/\nu=-6.3\times 10^{-7}$ [cf. Fig. 1 of \citet{grs15}] as a function of initial volume-averaged toroidal field strength $\langle B^i_t\rangle$, for $B_0=2.95\times 10^{13}$ G, $\mstar=1.4\msun$, and $\rstar=10^4$ m. The star has a quadrupole component, which decreases in magnitude post-anti-glitch. We keep $\sigma=-5$ and $\kappa_i=0.8$ (initial $\kappa$). The three curves represent different final $\kappa$ values: $\kappa_f=0.6$ (solid curve), $\kappa_f=0.4$ (dashed curve), and $\kappa_f=0.1$ (dotted curve). The thin solid curve is the identity function, for reference.}
 \label{fig3}
\end{figure}

In Fig. \ref{fig3}, we show $\langle B^f_t\rangle$ given by equation (10) for $\Delta\nu/\nu=-6.3\times 10^{-7}$ [cf. Fig. 1 of \citet{grs15}] as a function of $\langle B^i_t\rangle$, for $\mstar=1.4\msun$, $\rstar=10^4$ m, $B_0=2.95\times 10^{13}$ G, and three different $\kappa_f$ (final $\kappa$) values: $\kappa_f=0.6$ (solid curve), $\kappa_f=0.4$ (dashed curve), and $\kappa_f=0.1$ (dotted curve). We assume $\kappa_i=0.8$ (initial $\kappa$), i.e., that the initial surface quadrupole magnetic field strength at the pole is $1.6 B_0$ (surface dipole magnetic field strength at the pole is $2B_0$ as before). This rather extreme value is chosen for the sake of definiteness\footnote{Strong higher-order multipoles in magnetars, much stronger than the dipole component (assumed to be the component most responsible for spin down), are not ruled out completely. Analysis of the X-ray spectrum of SGR 0418$+$5729 suggests that the surface field strength is $\sim 10^2$ stronger than the spin-down-inferred field strength \citep{retal10,ggo11}.}. For the dipole-plus-quadrupole twisted torus, $\epsilon$ can be written as a function of $\langle B_t\rangle$ according to


\be \epsilon=a \left(\frac{B_0}{10^{13}\textrm{ G}}\right)^2 \left(\frac{\rstar}{10^4\textrm{ m}}\right)^4 \left(\frac{\mstar}{1.4\textrm{ }\msun}\right)^{-2} \left(1-\frac{b\langle B_t\rangle^2}{B_0^2}\right),\ee
where the values of $a$ and $b$ are given as functions of $\kappa$ in Table 1 (for $\kappa=0.1$, 0.2, 0.4, and 0.8; second and third columns in the table).

Figure \ref{fig3} tells us that, for a given $\langle B^i_t\rangle$, $\langle B^f_t\rangle$ decreases as $\kappa_f$ decreases. This is because, in general, the quadrupole component induces more positive (oblate) $\epsilon$: for a given $\langle B_t\rangle$, a configuration with $\kappa=0.1$ is more prolate (i.e., $\epsilon$ more negative) than one with $\kappa=0.6$, for example. Therefore, a decrease in $\kappa$ means a decrease in $\epsilon$, which must be countered by a decrease in $\langle B_t\rangle$ (which increases $\epsilon$) to obtain the $(\epsilon_i-\epsilon_f)$ value required to match the observed $\Delta\nu/\nu$ through equation (10). Compared to the pure dipole case of Section 2, however, we find that $\langle B_t^f\rangle$ is generally larger for the mixed case than for the pure dipole case. For example, with $\langle B_t^i\rangle=2\times 10^{15}$ G, one finds $\langle B_t^f\rangle=8.7\times 10^{14}$ G for the pure dipole case (see solid curve in Fig. 1), but $\langle B_t^f\rangle = 1.8\times 10^{15}$ G for the $\kappa_i=0.8$ to $\kappa_f=0.6$ transition (solid curve in Fig. 3). Note also that, while the general behaviour of the $\langle B_t^i\rangle$--$\langle B_t^f\rangle$ curves in Fig. 3 are similar to Fig. 1, the minimum values of $\langle B_t^i\rangle$ are smaller for the mixed cases, e.g., $\langle B_t^i\rangle\approx 3\times 10^{14}$ G for the transitions shown in Fig. 3, compared to $1.8\times 10^{15}$ G for the pure dipole case (Fig. 1, solid curve). The decay of the quadrupole plus toroidal components thus provides another channel for the anti-glitch to proceed, one which requires less initial toroidal field strength and less reduction in toroidal field strength, compared to the decay of only the toroidal component in a purely dipolar configuration.

It may be instructive to explore some other possible $\kappa$ transitions. Because the volume of the torus changes with $\kappa$ \citep{msm15}, we find it easier to express our configurations in terms of $\Lambda$, the ratio of the poloidal field energy to total field energy. In Fig. 4, we plot some combinations of $\kappa_f$ and $\Lambda_f$ (final $\Lambda$) which yield $\Delta\nu/\nu=-6.3\times 10^{-7}$ via equation (10), given initial values $\Lambda_i=0.1$ and $\kappa_i=1$ (triangles), 0.8 (pluses), and 0.6 (crosses). There is a clear trend that lower $\kappa_f$ (for each given $\kappa_i$) and higher $\kappa_i$ (for each given $\kappa_f$) need higher $\Lambda_f$ (i.e., a greater decrease of toroidal component). Figure 4 therefore shows more clearly how the poloidal quadrupole component tends to deform the star into a more oblate shape.

\begin{figure}
\centerline{\epsfxsize=13cm\epsfbox{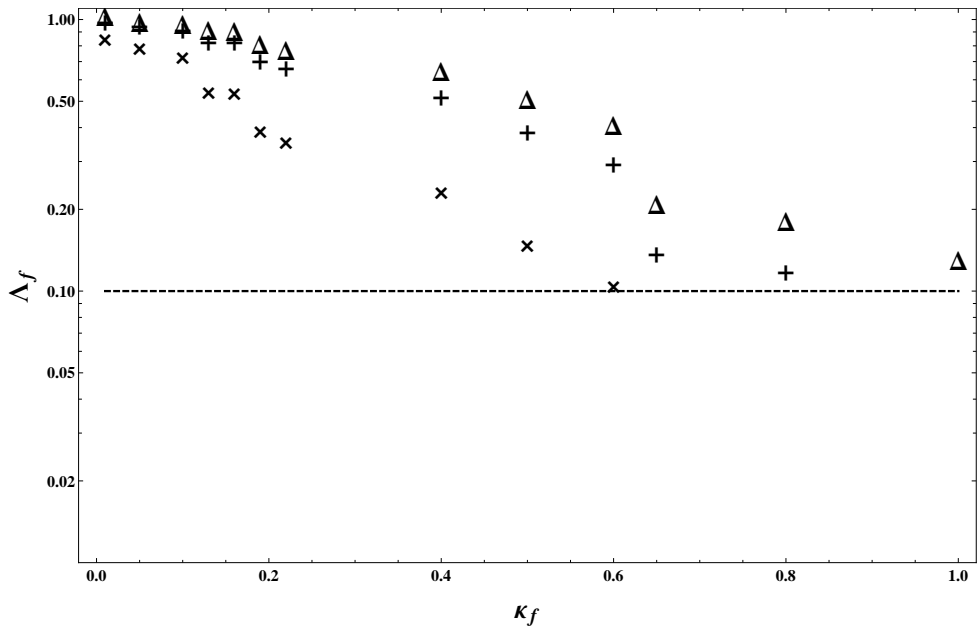}}
 \caption{Combinations of $\kappa_f$ and $\Lambda_f$ (final values of $\kappa$ and $\Lambda$) which produce the anti-glitch magnitude consistent with that observed in the AXP 1E 2259$+$586 ($\Delta\nu/\nu=-6.3\times 10^{-7}$) for a fixed $\Lambda_i$ (initial $\Lambda$). The star is in a dipole-quadrupole twisted torus configuration, with $\Lambda_i=0.1$ (this initial value of $\Lambda$ is indicated by the horizontal dashed line, for reference), and $\kappa_i=1$ (triangles), 0.8 (pluses), and 0.6 (crosses). The dipole surface polar field strength is kept constant at $5.9\times 10^{13}$ G (inferred from spin down) and we assume a canonical neutron star mass (1.4 $\msun$) and radius ($10^4$ m).}
 \label{fig4}
\end{figure}

\begin{table*}

 \begin{minipage}{145mm}
 \centering
  \caption{Values of the parameters $a$, $b$, $c$, and $d$ in the ellipticity and energy formulae, equations (14) and (16). The second and third columns give the values of $a$ and $b$ for different values of $\kappa$, given in the first column. The fourth and fifth columns give the values of $c$ and $d$ for transitions from $\kappa =0.8$ to the values of $\kappa$ given in the first column.}
  \begin{tabular}{@{}lcccc@{}}
  \hline
    $\kappa$ & $a$ ($10^{-8}$) & $b$ & $c$  & $d$\\
\hline
0.1 & $9.37\times 10^{-1}$  & $0.394$   &  4.90 & 0.63  \\
0.4 & $9.96\times 10^{-1}$  & $0.236$   & 4.24 & 0.73  \\
0.6 & $1.19$  & $0.149$   &  3.64 & 0.85  \\
0.8 & $1.77$  & $0.080$   &  -- & --  \\
\hline
\end{tabular}
\end{minipage}
\end{table*}

\section{Energetics}

In this section, we calculate the magnetic energy difference between the pre-anti-glitch and post-anti-glitch states and compare it to the observed energy output during the 2012 April 21 burst, which the anti-glitch accompanied \citep{aetal13b}. For the dipolar case discussed in Section 2, the change in magnetic energy $\Delta E_m$ can be written as a function of stellar radius and $\langle B_t\rangle$ as

\be \Delta E_m = 6.59\times 10^{46} \left(\frac{\rstar}{10^4\textrm{ m}}\right)^3 \left[\left(\frac{\langle B_t^f\rangle}{10^{15}\textrm{ G}}\right)^2 - \left(\frac{\langle B_t^i\rangle}{10^{15}\textrm{ G}}\right)^2\right] \textrm{ erg}.\ee
For the dipole-plus-quadrupole case described in Section 3, we find

\be \Delta E_{m} = c \times 10^{46} \left(\frac{\rstar}{10^4\textrm{ m}}\right)^3\left[ \left( \frac {\langle B_{t}^{f} \rangle} {10^{15} \text{G}} \right)^{2} - d \left( \frac {\langle B_{t}^{i} \rangle} {10^{15} \text{G}} \right)^{2}\right] \textrm{ erg},\ee
with the dipole surface polar field strength kept constant at $2.95\times 10^{13}$ G. The dimensionless factors $c$ and $d$ are given in Table 1 for three representative transitions from $\kappa_i=0.8$ to $\kappa_f=0.1$, 0.4, and 0.6.

During the anti-glitch, the \emph{Fermi} and \emph{Swift} satellites detected an energy release of $E\sim 10^{38}$ erg in the 10--$10^3$ keV band and $E\sim 10^{41}$ erg in the 2--10 keV band \citep{fetal12,aetal13b,hg14,grs15}. We obtain, for both the dipole and dipole-plus-quadrupole configurations, $\Delta E_m\sim 10^{45}$--$10^{47}$ erg, similar to the most energetic SGR giant flare \citep{petal05,m08}. If the anti-glitch was indeed due to a sudden $\epsilon$ change, this suggests two possibilities: (1) that, as proposed by \citet{grs15}, while the crust readjustment process (observed as the anti-glitch) is instantaneous [$\lesssim 1$ d \citep{aetal13b}], the magnetic field decay that leads to it is gradual, with time-scale $\sim 10^5$ yr; or (2) that the magnetic energy is released instantaneously, but is inefficiently ($\sim 10^{-5}$) converted into observed electromagnetic energy. As currently there has only been one confirmed anti-glitch, we cannot draw a conclusion yet.


\section{Anti-glitches in other magnetars?}

According to equation (9), we need $B_p\gtrsim 10^{13}$ G to get an anti-glitch of the size observed in AXP 1E 2259$+$586 with $\Lambda_i=10^{-3}$ [the lower bound for stability \citep{aetal13}] or with $\epsilon_\textrm{i}=-4\times 10^{-6}$ [the maximum ellipticity that the crust can tolerate before breaking \citep{hk09}]. Why, then, have we not observed an anti-glitch from other magnetars, many of which have higher $B_p$ than AXP 1E 2259$+$586? This is a question that requires a more detailed study and a more sophisticated model than the ones we discuss in this paper. The answer may simply be that we have not yet observed enough objects for long enough, or there may be a more microphysical explanation. Either way, using equations (9) and (10), we can set upper limits on the possible sizes of anti-glitches from other magnetars. Assuming that (1) the magnetic field is in a purely dipolar twisted torus configuration, (2) that the change in $\epsilon$ is caused entirely by a change of $\Lambda$ (or, equivalently, $\langle B_t\rangle$), like in Section 2, (3) that $\epsilon_\textrm{i}=-4\times 10^{-6}$ (the maximum $\epsilon$ that a neutron star's crust can support before cracking), (4) that $\epsilon_\textrm{f}=\epsilon_\textrm{max}$, and (5) $\mstar = 1.4$ $\msun$ and $\rstar=10^4$ m, we write down the maximum possible anti-glitch $(\Delta\nu/\nu)_\textrm{max}$ and the attendant magnetic energy change $(\Delta E_m)_\textrm{max}$ for four magnetars which have exhibited glitches (even though glitches and anti-glitches may be due to entirely different physical processes) and display them in Table 2.

As evident from Table 2, the maximum anti-glitches should be detectable easily by X-ray timing experiments similar to the ones targeting AXP 1E 2259$+$586. The stronger $B_p$ is, the larger the size of the maximum anti-glitch and the associated magnetic energy change.

The maximum anti-glitches in Table 2 are accompanied by large magnetic energy releases, stronger the one detected during AXP 1E 2259$+$586 anti-glitch, stronger than the 2004 December 27 giant flare of SGR 1806$-$20 ($\sim 10^{46}$ erg) \citep{petal05,m08}. If we assume the magnetic field change is instantaneous and that $\sim 10^{-5}$ of the released magnetic energy is converted into radiation as for AXP 1E 2259$+$586 (as discussed in Section 4), then we can expect the maximal anti-glitches from these sample magnetars to be accompanied by giant flares with energies $\sim 10^{44}$--$10^{45}$ erg. If these putative large anti-glitches are \emph{not} accompanied by energy releases of these magnitudes, we can conclude that, as proposed by \citet{grs15}, the magnetic fields must decay gradually before the crust cracks and readjusts, or that the electromagnetic conversion efficiency varies substantially across the population.

\begin{table*}

 \begin{minipage}{145mm}
 \centering
  \caption{Maximum anti-glitch magnitudes $(\Delta\nu/\nu)_\textrm{max}$ and the associated magnetic energy change $(\Delta E_m)_\textrm{max}$ for four glitching magnetars ($B_p$ taken from the McGill magnetar catalog), predicted by assuming that the magnetic field is in a purely dipolar twisted torus configuration, that the change in $\epsilon$ is caused entirely by a change of $\Lambda$, like in Section 2, and that one has $\epsilon=-4\times 10^{-6}$ initially [the maximum ellipticity that the crust can support before breaking \citep{hk09}], that $\epsilon_\textrm{f}=0$, and $\mstar = 1.4$ $\msun$ and $\rstar=10^4$ m.}
  \begin{tabular}{@{}lccc@{}}
  \hline
    Name & $B_p$ ($10^{14}$ G) & $(\Delta\nu/\nu)_\textrm{max}$ ($10^{-6}$) & $(\Delta E_m)_\textrm{max}$ ($10^{48}$ erg)\\
\hline
4U 0142$+$61 & 1.3 & $-2.7$ & $-9.2\times 10^{-1}$\\
1E 1048.1$-$5937 & 3.9 & $-3.0$ & $-1.0$\\
1RXS J170849.0$-$40091061 & 4.7 & $-3.2$ & $-1.1$\\
1E 1841$-$045 & 7.0 & $-3.9$ & $-1.3$ \\
\hline
\end{tabular}
\end{minipage}
\end{table*}

\section{Discussion}

In this paper, we refine the model proposed by \citet{grs15} to explain the anti-glitch of the AXP 1E 2259$+$586 in terms of a sudden change in ellipticity $\epsilon$. We also show how the observed anti-glitch can also occur via an alternative magnetic channel, namely the decay of a higher-order multipolar field (in addition to a decay of the toroidal component). We calculate the change in $\epsilon$ self-consistently for a dipole twisted torus in Section 2 and a dipole-quadrupole twisted torus in Section 3. We show that, by neglecting the contribution of the poloidal component to the Lorentz force, one overestimates the prolateness of the star (for example, for $1.9\times 10^{15}\textrm{ G}<\langle B_t\rangle<4.9\times 10^{15}\textrm{ G}$ and $B_p=5.9\times 10^{13}$ G, we find $\epsilon/\epsilon_\textrm{GRS}\sim 0.15$), which leads to an underestimate of the required $\langle B_t^i\rangle$ and the required change in field strength (Figs. 1 and 2). For both dipole and dipole-quadrupole twisted tori, we find that only objects with $B_p\gtrsim 10^{13}$ G (e.g., magnetars) can match the observed $\Delta\nu/\nu$, confirming a conclusion of \citet{grs15}. This means that we are not likely to observe a similar anti-glitch in a radio pulsar and that, despite superficial similarities, a magnetar anti-glitch may have an entirely different physical origin from a magnetar or pulsar glitch.

If the interior of a neutron star consists of superconducting protons \citep{l13} or quarks \citep{gjs12}, then $|\epsilon|$ is raised by a factor of $\sim H_{c1}/\langle B\rangle$, where $H_{c1}\sim 10^{16}$ G is the lower superconductivity critical field \citep{gas11}. In AXP 1E 2259$+$586, for example, a superconducting interior can raise $|\epsilon|$ by a factor of $\sim 10$, lowering the required $\langle B_t^i\rangle$ and $\Delta\langle B_t\rangle$ by a factor of $\sim 0.3$. Future anti-glitches, if observed in a star with a lower dipole field inferred from spin down, may be taken as evidence of superconducting interiors.

For $\epsilon$ to change by the amount needed to give the observed $\Delta\nu/\nu$ in AXP 1E 2259$+$586, the magnetic energy change is larger by 5 or 6 orders of magnitude than the observed outburst energy. In Section 5, we calculate the upper limits of anti-glitch magnitudes and energies in four other glitching magnetars (assuming for definiteness that they start at the edge of stability, $\Lambda=10^{-3}$). We find $(\Delta\nu/\nu)_\textrm{max}\sim 10^{-5}$--$10^{-3}$, higher than the observed AXP 1E 2259$+$586 anti-glitch, and magnetic energy releases of order $10^{49}$--$10^{50}$ erg, higher than the most energetic magnetar giant flare of $\sim 10^{46}$ erg \citep{petal05,m08}. If these anti-glitches are observed in the future, we can use the energy of the accompanying burst/flare (if any) to help conclude whether the field reconfiguration is instantaneous but with an inefficient radiative conversion or the field decay is a slow process \citep{grs15}. \citet{letal15} recently set an upper limit of $4\times 10^{46}$ erg on energy released by crustal fracture during a magnetar flare (largely independent of magnetic field strength), assuming the fracture extends to the base of the crust. Therefore, a high-energy flare can, in principle, be powered solely by a `crustquake'. Differentiating the energy contributions from crust and field may be difficult in practice, but if a flare is accompanied by an anti-glitch, a significant fraction of the observed energy may be due to field reconfiguration.



The anti-glitch in AXP 1E 2259$+$586 has also been interpreted as evidence that AXPs and other magnetars are surrounded by fallback matter \citep{k13}. \citet{hg14} proposed that the anti-glitch is due to a collision between AXP 1E 2259$+$586 and a small solid body with a mass $\sim 10^{18}$ kg. \citet{kg14} suggested instead that the anti-glitch is a consequence of neutron star superfluidity, just like a normal pulsar glitch, due to a velocity lag between the superfluid and the crust. We do not consider these alternative scenarios in this paper.

We also do not model the trigger and time-scale of the anti-glitch. Like \citet{grs15}, we implicitly assume that the anti-glitch (and the outburst it accompanied) occurred, when the magnetar crust cracked due to some internal instability, built up as the internal toroidal magnetic field decayed (but see also Sections 4 and 5). \citet{l14}, in contrast, concluded that magnetar bursts and flares must be caused by a relaxation of the \emph{external} field, to be consistent with the rise times of the accompanying quasi-periodic oscillations. \citet{l15} similarly concluded that a magnetar burst/flare can proceed without any internal instability leading to crust cracking. A simulation of the field decay process, the resultant crust cracking, and a calculation of the time-scales involved are reserved for future work.

\section*{Acknowledgments}

We thank the referee for the comments and suggestions, which have improved this manuscript considerably. This work was supported by an Australian Research Council Discovery Project Grant (DP110103347) and an Australian Postgraduate Award.

\bsp \label{lastpage}

\end{document}